\documentclass[%
 aip,
 amsmath,amssymb,
 reprint,%
]{revtex4-1}

\usepackage{graphicx}
\usepackage{dcolumn}
\usepackage{bm}

\usepackage[utf8]{inputenc}
\usepackage[T1]{fontenc}
\usepackage{mathptmx}
\usepackage{etoolbox}

\makeatletter
\def\@email#1#2{%
 \endgroup
 \patchcmd{\titleblock@produce}
  {\frontmatter@RRAPformat}
  {\frontmatter@RRAPformat{\produce@RRAP{*#1\href{mailto:#2}{#2}}}\frontmatter@RRAPformat}
  {}{}
}%
\makeatother
\begin{document}

\preprint{AIP/123-QED}

\title[Multi-Mode Input–Output Model for Cavity Magnonics]{Multi-Mode Input–Output Model for Cavity Magnonics: Phase-Resolved Control of Level Repulsion, Level Attraction, and Nonreciprocal Transmission}
\author{Guillaume Bourcin}
\affiliation{kwan-tek, 1 rue Galilée, Espace Innova, 56270 Ploemeur, France.}
\author{Mufti Avicena}
\affiliation{IMT Atlantique, Lab-STICC, UMR CNRS 6285, F-29238 Brest, France.}
\author{Vincent Vlaminck}
\affiliation{IMT Atlantique, Lab-STICC, UMR CNRS 6285, F-29238 Brest, France.}
\author{Jeremy Bourhill}
\affiliation{Quantum Technologies and Dark Matter Labs, Department of Physics,
University of Western Australia, 35 Stirling Hwy, 6009 Crawley, Western Australia.}
\author{Vincent Castel}
\affiliation{IMT Atlantique, Lab-STICC, UMR CNRS 6285, F-29238 Brest, France.}
 \email{vincent.castel@imt-atlantique.fr}
\date{\today}

\begin{abstract}
We experimentally validate a unified input--output model that incorporates internal and external coupling phases across multiple cavity modes in a room-temperature cavity magnonic system. By explicitly accounting for both phase contributions, the model provides a clear interpretation of the transition from level repulsion to level attraction at an interference-induced antiresonance, and accurately reproduces nonreciprocal transmission arising from the internal phases of the contributing modes. Quantitative agreement between experiments and simulations is obtained across all coupling regimes, establishing a predictive framework for phase-controlled cavity--magnon devices including isolators, circulators, and quantum transducers.
\end{abstract}

\maketitle

\begin{quotation}
We present a multi-mode input--output model that captures the transition between attractive and repulsive coupling regimes by explicitly accounting for internal and external coupling phases. Validated by experiments and numerical simulations, the model provides a unified description of coupling dynamics featuring interference-induced antiresonances and nonreciprocal transmission, and offers clear design guidelines for tunable nonreciprocal devices.
\end{quotation}

\section{\label{sec:level1}Introduction}

Cavity magnonics provides a powerful platform for studying coherent interactions between collective spin excitations (magnons) and microwave photons, and for exploring light–matter hybridization in the strong- and ultrastrong-coupling regimes\cite{Huebl2013,Tabuchi2014}. A key aspect of magnon–photon interactions is the transition between level repulsion and level attraction, which reflects fundamentally different coupling mechanisms between the hybridized modes. Understanding and controlling this transition is essential for engineering reconfigurable magnon–photon devices and for advancing cavity-based signal processing architectures. An early demonstration by Rao et al.~\cite{rao_2019} reported the transition between level attraction and level repulsion at an interference-induced antiresonance (referred to simply as antiresonance in the following) in a three-dimensional cavity--magnon system, providing a foundational experimental reference for the study of antiresonances in cavity magnonics. This transition has also been observed in two-dimensional circuit geometries as a function of frequency~\cite{PhysRevLett.128.047701}. The antiresonance is not a cavity mode; it is defined as a frequency at which destructive interference between cavity modes --- governed by the external coupling phases and external dissipation rates --- produces a pronounced dip in the transmission~\cite{Bourcinattraction}. Building on the work of Rao et al., we reconstructed that system numerically, identified the parameters controlling the antiresonance position, and developed the theoretical framework introduced in our previous publication~\cite{Bourcin2024}. The present work constitutes a direct experimental and theoretical extension of that line of research, incorporating a multi-mode internal phase picture.

Previous studies have shown that nonreciprocal transmission can naturally emerge from the underlying coupling dynamics between magnons and photons. Wang et al.~\cite{wang2019} demonstrated isolation ratios exceeding 20~dB in an open cavity--magnon system. More recent works have further explored nonreciprocity through interference mechanisms~\cite{Rao2021}, feedback circuits~\cite{wangRealizationUnidirectionalAmplification2023}, irregular resonant cavities~\cite{zhang2021}, cryogenic setups~\cite{kim2024}, while theoretical models have investigated related effects arising from Kerr nonlinearities~\cite{wu2024} and magnomechanical interactions~\cite{wuNonreciprocalMechanicalSqueezing2024}.

Despite these advances, existing input--output models often neglect the role of coupling phases, which are crucial for accurately describing interference effects in multi-mode cavity--magnon systems. Two distinct phases can be identified: (i)~the internal coupling phase, or YIG-experienced phase, which reflects the polarization state of the RF magnetic field of each individual cavity mode at the YIG location and depends on the cavity mode structure and sphere placement~\cite{Gardin2023, Gardin2024}; and (ii)~the external coupling phase, or probe-field phase, which originates from the phase delay associated with the excitation and detection ports and determines how the incoming microwave probe interferes with the intracavity field~\cite{Bourcin2024}. Importantly, these two phase contributions govern distinct physical observables: external phases determine the character of the antiresonance, while internal phases govern the nonreciprocal transmission. Although achieving nonreciprocity through the local microwave polarization at the YIG position is a known concept in single-mode cavity magnonics, the multi-mode situation is substantially more complex. In particular, the relevant quantity is not the field polarization at the antiresonance frequency itself, but rather the polarization state of each individual cavity mode evaluated at the YIG sphere. The contributions of these modes can compete constructively or destructively, and identifying which modes enhance or suppress nonreciprocity is essential for engineering large isolation ratios.

In this work, we develop an input--output model that explicitly incorporates both internal and external coupling phases across multiple cavity modes. The model provides a unified description of phase-dependent level repulsion and level attraction at an interference-induced antiresonance. We show that the model accurately predicts the experimentally observed transition between repulsive and attractive coupling regimes in a three-dimensional cavity coupled to a 1-mm YIG sphere, and simultaneously reproduces the emergence of nonreciprocal transmission arising from the internal phases of the contributing modes. Our combined theoretical, numerical, and experimental approach provides a consistent and predictive framework for quantifying and controlling hybrid magnon--photon coupling. 

\section{\label{sec:level1}Cavity magnonics system}

\subsection{\label{sec:level2}Experimental details}
The cavity magnonics system is composed of a 1 mm diameter pure Yttrium Iron Garnet (YIG) sphere and a double-post re-entrant cavity,
previously described in Ref. \cite{PhysRevApplied.2.054002}. The cavity\cite{Bourcin2023} includes two access ports, enabling the measurement of transmission and reflection parameters (magnitude and phase) at an input power of -10 dBm between 10 and 13 GHz using a vector network analyzer (VNA, R$\&$S ZNB 43.5 GHz) at room temperature. Coaxial loop antennas, which couple to the azimuthal component of the magnetic field, are positioned at these ports to both drive and read out microwave signals. The YIG sphere is positioned at various locations inside the cavity. To ensure precise and reproducible placement, we machined a 1 mm thick Rohacell foam sample holder. Due to its very low dielectric constant and loss tangent, as well as its location inside the cavity—far from the regions of highest RF electric field concentrated between the post tips and the cavity lid—the Rohacell holder does not measurably affect the cavity response. The holder was specifically designed to fit tightly between the inner edges of the cavity and the re-entrant posts, providing both mechanical stability and repeatable alignment of the YIG sphere.

To achieve an antiresonance at a sufficiently low frequency, an alumina slab with relative permittivity $\varepsilon_\mathrm{r} = 10$ and dimensions of 6.2 × 3.8 × 0.56 mm$^3$ was inserted between the two posts, similarly to the YIG slab placement in Ref.\cite{Bourcin2023}. A numerical parametric study of the dielectric dimensions was conducted to tailor the cavity response, enabling the antiresonance to lie between two adjacent modes with a high quality factor (as shown in Appendix \ref{appendix:antires}). The measured transmission of the optimized dielectric-loaded cavity without a YIG sphere is shown in Fig. \ref{fig:FW_FIG2}(a) as a solid black line, where the antiresonance frequency of interest is 11.46 GHz. This configuration serves as the reference for the subsequent investigation of the magnon–photon coupling regimes and the comparison with the input–output model.

\subsection{\label{sec:level2}Input--output theory}
The internal coupling phase~\cite{Gardin2023,Gardin2024} is determined by the spatial orientation of the RF magnetic field that excites the YIG sphere (the YIG-experienced phase) and is defined as
\begin{equation}
\label{internalphase}
    \theta_{uv}
    = \arg \left(
        \int_{V_{m_v}} H_x^u(\mathbf{r})\, d^3 r
        + i \int_{V_{m_v}} H_y^u(\mathbf{r})\, d^3 r
    \right),
\end{equation}
where \(H_x\) and \(H_y\) are the orthogonal components of the cavity RF magnetic field within the magnetic sample volume \(V_{m_v}\), with the static applied field aligned along the (z)-axis. Physically, \(\theta_{uv}\) quantifies the internal phase accumulated by the YIG through its interaction with the cavity field and directly reflects the polarization state of the RF field inside the sample. The index \(u\) runs over the four cavity modes, and \(v\)=0 labels the unique magnon mode. The corresponding coupling strength is written as \(g_{u0}\, e^{i\theta_{u0}}\), where $g_{u0}$ is the coupling strength magnitude between cavity photon u and the magnon.

The external coupling phase~\cite{Bourcin2024} is defined by the direction of the RF magnetic field of the excited cavity mode as sensed by the physical loop ports of the cavity (port-field phase) and is denoted by \(\phi_{up}\), where $p$ denotes the port number. Within the Markov approximation~\cite{markov}, the external coupling strength between the cavity modes and the
ports is assumed to be frequency-independent and can therefore be written as
\begin{equation}
    \kappa_{up}(\omega) \;\Rightarrow\;
    \kappa_{up} = \sqrt{\gamma_{up}^\mathrm{ext}}\, e^{i\phi_{up}},
\end{equation}
where $\gamma_{up}^\mathrm{ext}$ denotes the external dissipation rate. In our setup, the three-dimensional microwave cavity is coupled to external ports via handmade inductive loops based on $50~\Omega$ coaxial cables. While the inductive coupling formally leads to an ohmic spectral density, its variation over the cavity linewidth is negligible. The external electromagnetic bath can therefore be treated as effectively flat, justifying the use of the Markov approximation~\cite{markov}. 

In the cavity configuration with two input/output ports ($p \in \{0,1\}$), the coupling is implemented via loop probes located on opposite sides of the cavity, aligned along the azimuthal direction. As a result, the relative phase accumulated between the two ports can take only two possible values, $0$ or $\pi$. This phase depends on whether the resonant cavity mode is symmetric or antisymmetric with respect to the cavity mid-plane, respectively.

We model the system by considering a total Hamiltonian comprising 5 internal bosonic modes (4 photon modes and 1 magnon mode) coupled to 2 bath modes (input/output ports). The total Hamiltonian is written as $H = H_{\text{int}} + H_{\text{ext}}$ where $H_{\text{int}}$ is the Hamiltonian of the internal bosonic modes following the second quantization formalism \cite{dirac_principles_1981} and its internal coupling between the cavity modes and the magnon mode quantified by $g_{u0}$. We define $H_{\text{int}}$ under the rotating wave approximation\cite{RWA} as 

\begin{equation}
    \begin{split}
          H_{\text{int}}/\hbar &=
        \sum_{u=0}^{3} \omega_{c_u}\, \hat{c}_u^\dagger \hat{c}_u
        + \omega_{m}\, \hat{m}^\dagger \hat{m} \\
        &\quad + \frac{1}{2} \sum_{u=0}^{3}
        \left(
            g_{u0}\, \hat{c}_u \hat{m}^\dagger
            + g_{u0}^{*}\, \hat{c}_u^\dagger \hat{m}
        \right),
    \end{split}
\end{equation}
where $\hat{c}_u$ ($\hat{c}_u^\dagger$) is the bosonic annihilation (creation) operator of cavity mode $u \in \{0,1,2,3\}$, satisfying the commutation relation $[\hat{c}_u, \hat{c}_{u'}^\dagger] = \delta_{uu'}=1$. Similarly, $\hat{m}$ ($\hat{m}^\dagger$) denotes the annihilation (creation) operator of the single magnon mode, satisfying $[\hat{m}, \hat{m}^\dagger] = 1$.

Next, we have the harmonic oscillator bath modes describing the probes and their interactions with the cavity modes, expressed by the external Hamiltonian as follows:

\begin{equation}
    \begin{split}
        H_{\text{ext}}/\hbar = \int_\mathbb{R} &\omega \sum_{p=0}^1  \, (b^\dagger_p b_p) \\ & +\frac{i}{\sqrt{2 \pi}} \int_\mathbb{R} \omega \sum_{u  = 0}^3 \sum_{p=0}^1 \, (\kappa_{up} b^\dagger_{p}c_u  - h.c),
    \end{split}
\end{equation}
where $h.c$ represents the hermitian conjugate. The first term of this Hamiltonian represents the photon bath of each port as a continuous ensemble of harmonic oscillators in the frequency space, and the second one is the coupling between the photon bath of each port and the cavity photons. Based on the total system Hamiltonian, the \(2\times 2\) scattering (S-)matrix for the two ports is obtained from standard input–output theory and it can be written as\cite{FanSeq}:
\begin{equation}
\label{Spara}
    \mathbf{S}(\omega) = \mathbf{C} + \mathbf{D}\,\bigl[ -i\omega\,\mathbf{I} - \mathbf{A} \bigr]^{-1}\,\mathbf{B},
\end{equation}
where \(\mathbf{A}\) is the internal-mode matrix, \(\mathbf{B}\) is the coupling matrix between the two-port baths and the internal modes, and \(\mathbf{I}\) is the identity matrix. \(\mathbf{C}\) accounts for background, non-resonant scattering that directly connects the ports (crosstalk), independently of the internal resonant dynamics. For compactness of the matrix expression, we define $\tilde{\omega}_{cu}= \omega_{cu} -i \gamma_u^\mathrm{int}/2$ with $\gamma_u^\mathrm{int}$ is the internal dissipation rate and 
$\tilde{\omega}_{m}= \omega_{m} -i k_{m0}/2$ with $k_{m0}$ the dissipation of magnon 0 mode.

\[
\mathbf{A} =
\begin{pmatrix}
 -i\tilde{\omega}_{c0} & 0 & 0 & 0 & i g_{00} e^{i\theta_{00}} \\
 0 & -i\tilde{\omega}_{c1} & 0 & 0 & i g_{10} e^{i\theta_{10}} \\
 0 & 0 & -i\tilde{\omega}_{c2} & 0 & i g_{20} e^{i\theta_{20}} \\
 0 & 0 & 0 & -i\tilde{\omega}_{c3} & i g_{30} e^{i\theta_{30}} \\
 i g_{00} e^{-i\theta_{00}} & i g_{10} e^{-i\theta_{10}} & i g_{20} e^{-i\theta_{20}} & i g_{30} e^{-i\theta_{30}} & -i\tilde{\omega}_{m}
\end{pmatrix}
\]
\[
\mathbf{B} =
\begin{pmatrix}
\sqrt{\gamma_{00}^\mathrm{ext}}\, e^{i \phi_{00}} & \sqrt{\gamma_{01}^\mathrm{ext}}\, e^{i \phi_{01}} \\
\sqrt{\gamma_{10}^\mathrm{ext}}\, e^{i \phi_{10}} & \sqrt{\gamma_{11}^\mathrm{ext}}\, e^{i \phi_{11}} \\
\sqrt{\gamma_{20}^\mathrm{ext}}\, e^{i \phi_{20}} & \sqrt{\gamma_{21}^\mathrm{ext}}\, e^{i \phi_{21}} \\
\sqrt{\gamma_{30}^\mathrm{ext}}\, e^{i \phi_{30}} & \sqrt{\gamma_{31}^\mathrm{ext}}\, e^{i \phi_{31}} \\
0 & 0
\end{pmatrix}
\]
\[
\mathbf{C} =
\begin{pmatrix}
\sqrt{1-\xi} & \sqrt{\xi} \\ 
\sqrt{\xi} & \sqrt{1 - \xi}
\end{pmatrix}
\]
\[
\mathbf{D} = -\mathbf{CB}^\dagger, \\
\]
Since we assume no crosstalk, we set $\xi = 0$. The S-parameter response can then be numerically calculated as a function of frequency and applied magnetic field. 

To reproduce the transmission trace obtained from the measurement of the empty cavity (without YIG sphere) at room temperature shown in Fig. \ref{fig:FW_FIG2}(a), four cavity modes were included in the input–output model. The external dissipation rates $\gamma_{up}^\mathrm{ext}$ and intrinsic dissipation rates $\gamma_u^\mathrm{int}$ were extracted by fitting the measured transmission data. The fitting was performed using Eq.~(\ref{Spara}), including four cavity modes while excluding the magnon contribution. The resulting computed transmission, shown as a dashed red line in Fig. \ref{fig:FW_FIG2}(a), reproduces the experimental data with excellent agreement over the frequency range from 3.3 to 15.7 GHz and represented by the two grey vertical dashed lines. The corresponding fitted parameter values are summarized in Tab. \ref{tab:fitted_gammas}.

\begin{table} [h!]
\caption{\label{tab:fitted_gammas}Fitted values of the external dissipation rates $\gamma_{ip}^\mathrm{ext}$ with $\gamma_{u0}^\mathrm{ext}=\gamma_{u1}^\mathrm{ext}$ and the internal dissipation rates $\gamma_{u}^\mathrm{int}$ for the first four modes.}
\begin{ruledtabular}
\begin{tabular}{lcc}
Modes & $\bm{\gamma_{u0}^\mathrm{ext}=\gamma_{u1}^\mathrm{ext}}$ [$2\pi\times$MHz] & $\bm{\gamma_u^\mathrm{int}}$ [$2\pi\times$MHz]\\
\hline
    TM$_{\bm{010}}$ & $1.186 \pm 0.006$ & $3.4 \pm 0.7$\\
    TM$_{\bm{110}}$ & $0.2334 \pm 0.0007$ & $20.2 \pm 0.7$\\
    TM$_{\bm{210}}$ & $7.56 \pm 0.03$ & $38.1 \pm 1.3$\\
    TM$_{\bm{310}}$ & $8.89 \pm 0.04$ & $40.2 \pm 1.7$\\
\end{tabular}
\end{ruledtabular}
\label{modelist}
\end{table}

Owing to the mirror-symmetric construction of the two inductive loop probes and their identical geometrical and electrical properties, the external coupling rates to the cavity mode are assumed to be equal, $\gamma_{u0}^\mathrm{ext}=\gamma_{u1}^\mathrm{ext}$. Parameter uncertainties in Tab. \ref{tab:fitted_gammas} are estimated from the covariance matrix associated with the nonlinear least-squares fitting model. The covariance matrix is provided by the fitting routine implemented in Python and reflects the statistical uncertainties of the fitted parameters within the assumed model.


\subsection{\label{sec:level2}Finite element simulation}
\begin{figure} [h!]
\centering
\includegraphics[width=8.5 cm]{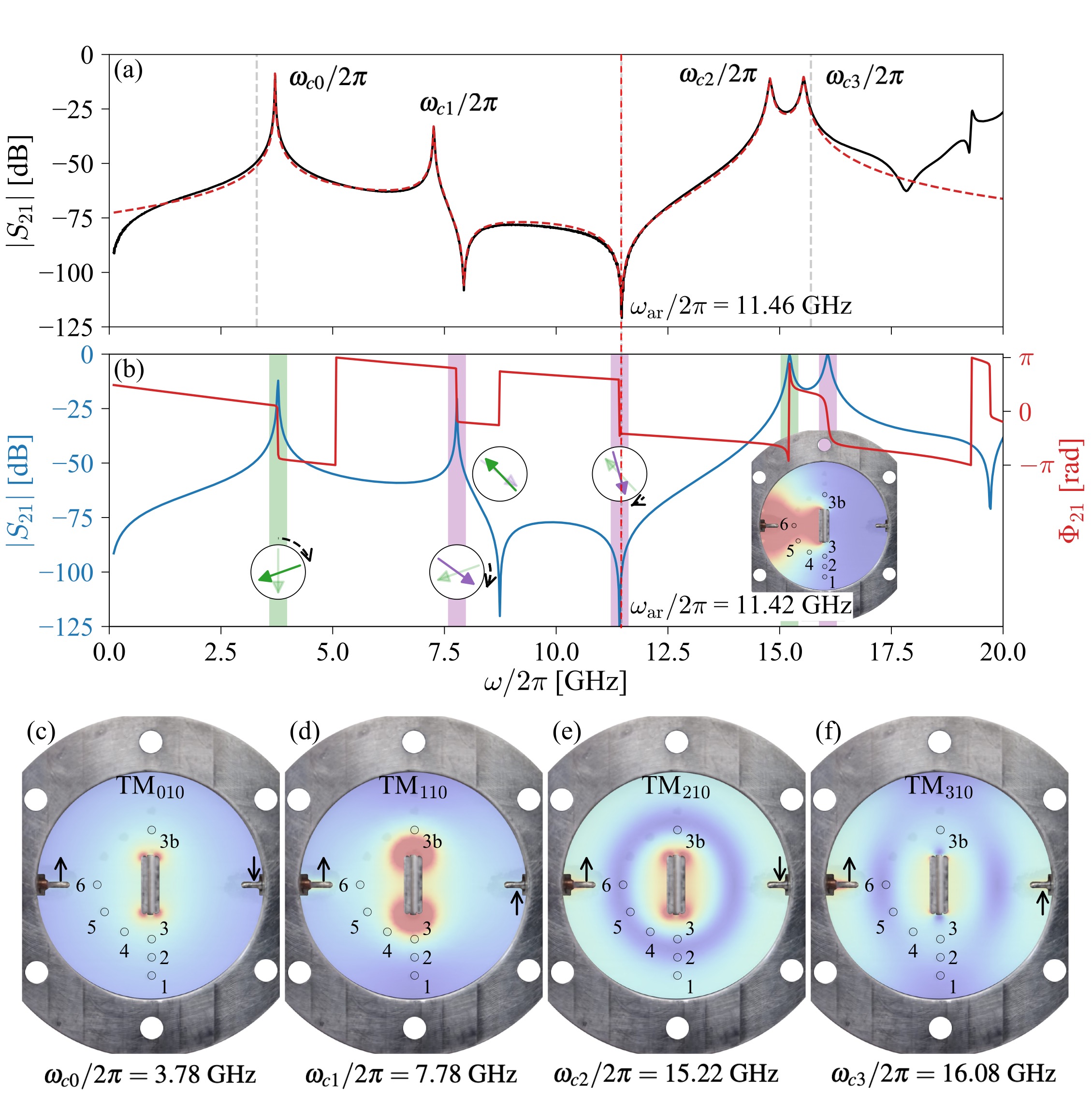}
\caption{(a) Transmission of the dielectric-loaded cavity: Measurement in solid black line; and input-output model (between 3.3 to 15.7 GHz, with fit region indicated by dashed grey lines) with fitted external and intrinsic dissipation rates values in dashed red line. The measurement was conducted without a YIG sphere inside the cavity. (b) Frequency domain simulation of the transmission of the dielectric-loaded cavity with a dielectric width of 566 $\mu$m. The magnitude is depicted in blue, and the phase in red. The modes presenting the same $\Phi_{21}$ jump as the antiresonance at 11.46 GHz are illustrated in purple areas, while the modes presenting an opposite-phase as the antiresonance are illustrated in green areas. The dotted vertical red line indicates the measured antiresonance frequency. (c)-(f) depict the RF magnetic fields of the first four modes and the different YIG positions. Black arrows at the probe locations indicate the orientation of the RF field.}
\label{fig:FW_FIG2}
\end{figure}

In Fig.~\ref{fig:FW_FIG2}(b), the simulated transmission (without the YIG sphere) is shown as a solid blue line for a dielectric loaded between the posts. 
Finite element simulations were performed using COMSOL Multiphysics. While the simulated cavity-mode frequencies do not exactly match the measured values, the antiresonance of interest at 11.42 GHz lies very close to the experimental one at 11.46 GHz. The phase of $S_{21}$, $\Phi_{21}$, plotted as a solid red line, shows that the first and third modes ($\omega_{c0}$ and $\omega_{c2}$) exhibit $\Phi_{21}$ jumps of the same sign, highlighted by the green shaded regions. In contrast, the second and fourth modes ($\omega_{c1}$ and $\omega_{c3}$), indicated by the purple shaded regions, display $\Phi_{21}$ jumps of opposite sign relative to $\omega_{c0}$ and $\omega_{c2}$.

Following Refs.~\cite{Bourcinattraction, Labource}, these phase jumps are categorized into two types: jumps from $0$ to $\pi$ and jumps from $\pi$ to $0$, referred to as \textit{left} and \textit{right} phase jumps, respectively, based on their direction on the trigonometric circle, regardless of their clockwise or counterclockwise orientation. It has been analytically demonstrated in Ref.~\cite{Bourcinattraction} that these two types of phase jumps play distinct roles in the effective coupling between the magnon mode at the antiresonance: a cavity mode exhibiting an identical (opposite) phase jump to that of the antiresonance results in an attractive (repulsive) effective coupling. Consequently, the nature of the effective coupling depends on the interplay between the phase jumps and the coupling strengths of the cavity modes with the magnon.

The antiresonance at $\omega_{ar}/2\pi = 11.42~\mathrm{GHz}$ exhibits the same $\Phi_{21}$ jump as cavity modes $\omega_{c1}$ and $\omega_{c3}$, indicating that their respective coupling strengths to the YIG sphere ($g_{10}/2\pi$ and $g_{30}/2\pi$) govern the attractive signature observed at the antiresonance, while the two remaining modes are responsible for the repulsive signature. Hereafter, the terms \textit{attraction} and \textit{repulsion} always refer to their effect on the antiresonance at $\omega_{ar}$. Notably, even in the absence of a YIG sphere, the attractive or repulsive coupling regime at $\omega_{ar}$ can be predicted solely from the analysis of the $\Phi_{21}$ phase jumps.

The predefined positions of the sample holder define six possible YIG-sphere placements, numbered from 1 to 6 in the inset (and Fig. \ref{fig:FW_FIG2}(c)-(f)). An additional position, denoted 3b, is introduced for experiments involving two YIG spheres (positions 3 and 3b), as described in Appendix~\ref{appendix:YIGposmeas}. Figures \ref{fig:FW_FIG2}(c)–(f) display the first four cavity modes: TM$_{\bm{010}}$ at $\omega_{c0}/2\pi = 3.78~\mathrm{GHz}$, TM$_{\bm{110}}$ at $\omega_{c1}/2\pi = 7.78~\mathrm{GHz}$, TM$_{\bm{210}}$ at $\omega_{c2}/2\pi = 15.22~\mathrm{GHz}$, and TM$_{\bm{310}}$ at $\omega_{c3}/2\pi = 16.08~\mathrm{GHz}$. Inspection of the field distributions reveals distinct spatial dependences. At position~1 (Pos.~1), the two attractive modes ($\omega_{c1}$ and $\omega_{c3}$) exhibit minimal field intensity, whereas their fields are significantly stronger at position~3 (Pos.~3). In contrast, the repulsive modes ($\omega_{c0}$ and $\omega_{c2}$) show relatively constant field intensities between Pos.~1 and Pos.~3. It should be noted that the field distributions in Figs. \ref{fig:FW_FIG2}(c)–(f) are not normalized, and their color scales were saturated to emphasize spatial variations. The black arrows at the probe locations in Figs.~\ref{fig:FW_FIG2}(c)-(f) indicate the orientation of the RF magnetic field evaluated at the center of each loop at $t=0$, from which the port-field phase $\phi_{up}$ at each probe location is directly obtained. As discussed previously, the phase difference between ports~1 and~2 can only take the values $0$ ($\omega_{c1}$ and $\omega_{c3}$) or $\pi$ ($\omega_{c0}$ and $\omega_{c2}$) for this cavity configuration. With a sufficient number of cavity modes, the external phases established here explain the experimentally observed antiresonance and enable assessment of each mode's contribution to attractive or repulsive coupling at the antiresonance.

\section{\label{sec:level1}Results and discussion}

\begin{figure} [h!]
\centering
\includegraphics[width=9 cm]{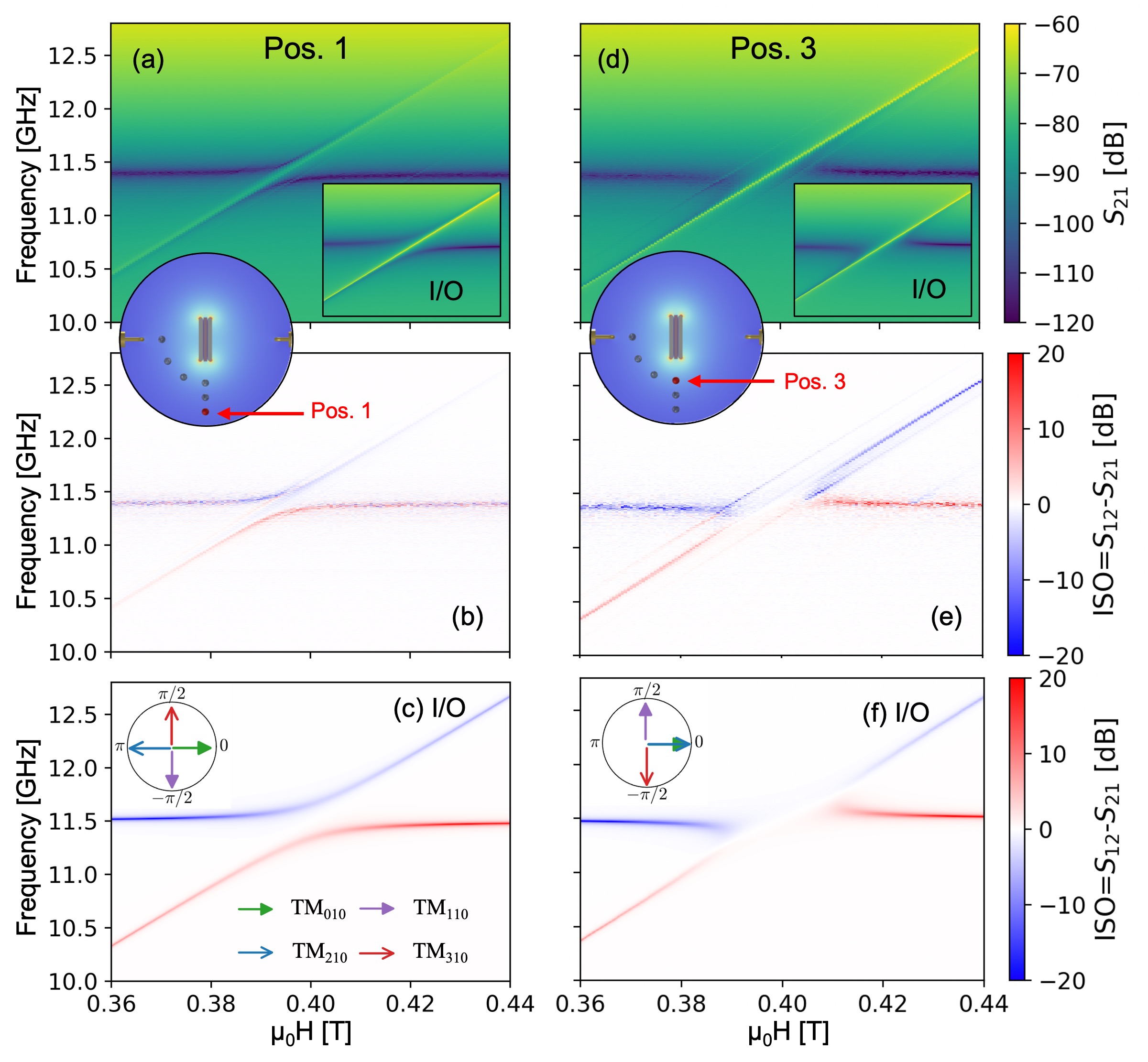}
\caption{Left panel -- repulsive coupling regime: (a) Measured transmission magnitude $S_{21}$, with the inset displaying the corresponding $S_{21}$ calculated using Eq.~\ref{Spara} and the parameters summarized in Tab.~\ref{tab:coupling_res}. The inset between (a) and (b) illustrates the position of the YIG sphere in the cavity. (b) and (c) show the measured and calculated isolation (ISO) parameters, defined as the difference between $S_{21}$ and $S_{12}$, respectively. The inset in (c) shows the internal coupling of each modes : TM$_{\bm{010}}$, TM$_{\bm{210}}$, TM$_{\bm{110}}$ and TM$_{\bm{310}}$. Right panel -- attractive coupling regime: (d)–(f) display the same measurements and calculations for Pos.~3, where the coupling becomes attractive.}
\label{fig:pos13}
\end{figure}

Having established the correspondence between the measured transmission spectra and the simulated cavity modes (without YIG), we now analyze how the coupling regime evolves as the YIG sphere is displaced within the cavity. Although several sphere positions were measured, many exhibit similar behavior (see Appendix~\ref{appendix:YIGposmeas}). To focus the discussion on the key physical effects, we select two representative positions that illustrate contrasting coupling regimes: one exhibiting predominantly repulsive coupling (Fig.~\ref{fig:pos13} (a)-(c)) and the other predominantly attractive coupling (Fig.~\ref{fig:pos13} (d)-(f)). 

The coupling strengths for each cavity mode were extracted from eigenmodes obtained via finite-element simulations, following the procedure developed in Refs.~\cite{bourhillUniversalCharacterizationCavity2020,Bourcinattraction}. From the measured spectra, the gyromagnetic ratio was estimated as $\gamma = 28.74 \pm 0.20$~GHz$\cdot$T$^{-1}$, and the Gilbert damping parameter set to $\alpha = 4\times10^{-4}$ to reproduce the experimental observations. This increase in damping relative to the intrinsic YIG linewidth commonly reported in the literature may arise from the simultaneous excitation of multiple magnon modes by the inhomogeneous RF field.~\cite{PhysRevB.96.064407}. Based on $\alpha$, the magnon decay rate is set to $k_{m0}/2\pi = 10$~MHz for a resonance at 11.5~GHz. For each mode $\omega_{cu}$, the internal phase $\theta_{u0}$ is determined by solving Eq.~\ref{internalphase}, which involves integrating the field components $H_x(\mathbf{r})$ and $H_y(\mathbf{r})$ over the YIG volume $V_m$ using COMSOL Multiphysics. The resulting internal phases are illustrated in the insets of Fig.~\ref{fig:pos13}(c) and (f). The external phase, $\phi_{up}$, is the phase delay determined by the direction of the RF field at the input/output loop centers, as discussed above. All parameters used in the input–output model are summarized in Tab.~\ref{tab:coupling_res} (see Appendix~\ref{appendix:EMandcoupling}). In the following, we analyze the attractive and repulsive coupling regimes separately, highlighting the spectral and phase characteristics that govern the observed nonreciprocal behavior.

Figure~\ref{fig:pos13} illustrates how the coupling regime can be tuned by varying the position of the YIG sphere inside the cavity. When the sphere is placed at Pos.~1, corresponding to a repulsive coupling configuration, the isolation spectra shown in Fig.~\ref{fig:pos13}(c) display a pronounced nonreciprocal response. In contrast, moving the sphere to Pos.~3 switches the system to an attractive coupling regime, leading to qualitatively different spectral features, as shown in Fig.~\ref{fig:pos13}(d)–(f). Figures~\ref{fig:pos13}(a) and (d) display the measured transmission magnitude $S_{21}$ for the two configurations. For each case, an inset shows the transmission calculated from the input–output model described above, using identical frequency and magnetic-field scales. The corresponding isolation maps in Fig.~\ref{fig:pos13}(b) and (e) directly reveal nonreciprocity in both coupling regimes. In our cavity configuration, the model predicts no nonreciprocity when internal phases are
excluded. However, because the antiresonance considered here arises from interference, the external phases cannot be tuned independently to assess their contribution to the nonreciprocal response. A clearer understanding of the respective roles of the number of modes and of the internal and external phases would likely require a different configuration, for instance one based on the standard coupling between a cavity mode and a magnon. Importantly, the calculated isolation captures not only the presence of nonreciprocal behavior but also the sign, magnitude, and frequency dependence of the isolation peaks observed experimentally. 

Some discrepancies between the measured and calculated spectra can nevertheless be identified. In particular, the antiresonance appears at slightly higher frequencies in the theoretical curves than in the experimental data. This difference originates from the interplay between the various cavity modes: even a small frequency shift of one or several modes is sufficient to noticeably displace the resulting antiresonance frequency. This sensitivity is directly visible in Figs.~\ref{fig:pos13}(b) and (c), where the experimental antiresonance occurs below 11.5~GHz, whereas it is predicted at higher frequencies in the calculated spectra. As a consequence, the YIG resonance intersects the antiresonance at slightly different magnetic-field values in experiment and in simulation. In addition, besides the uniform Kittel mode, couplings to other resonances are present and not fully captured by the model. These extra features observed at Pos. 3 are due to higher-order magnetostatic volume modes of the YIG sphere. Finally, the simulated coupling strength is slightly overestimated at Pos. 1, leading to a larger level repulsion in the calculated spectra. These effects explain most of the residual discrepancies between theory and experiment. 

It is worth noting that the input–output theory used here does not involve fitting to the measured spectra, except for the initial determination of the internal and external dissipation rates ($\gamma_u^\mathrm{int}$ and $\gamma_{up}^\mathrm{ext}$) and the gyromagnetic ratio. Specifically, the dissipation rates were obtained by fitting the measured transmission of the empty cavity (without YIG), while all other parameters were extracted independently from finite-element simulations, as described above. In this way, the model used to calculate the spectra contains no adjustable parameters. The input quantities include the cavity-mode frequencies (4), the coupling strengths (4), the external and internal coupling phases (8 + 4), and the magnon dissipation (1), for a total of 21 parameters.

\begin{figure} [h]
\centering
\includegraphics[width=8 cm]{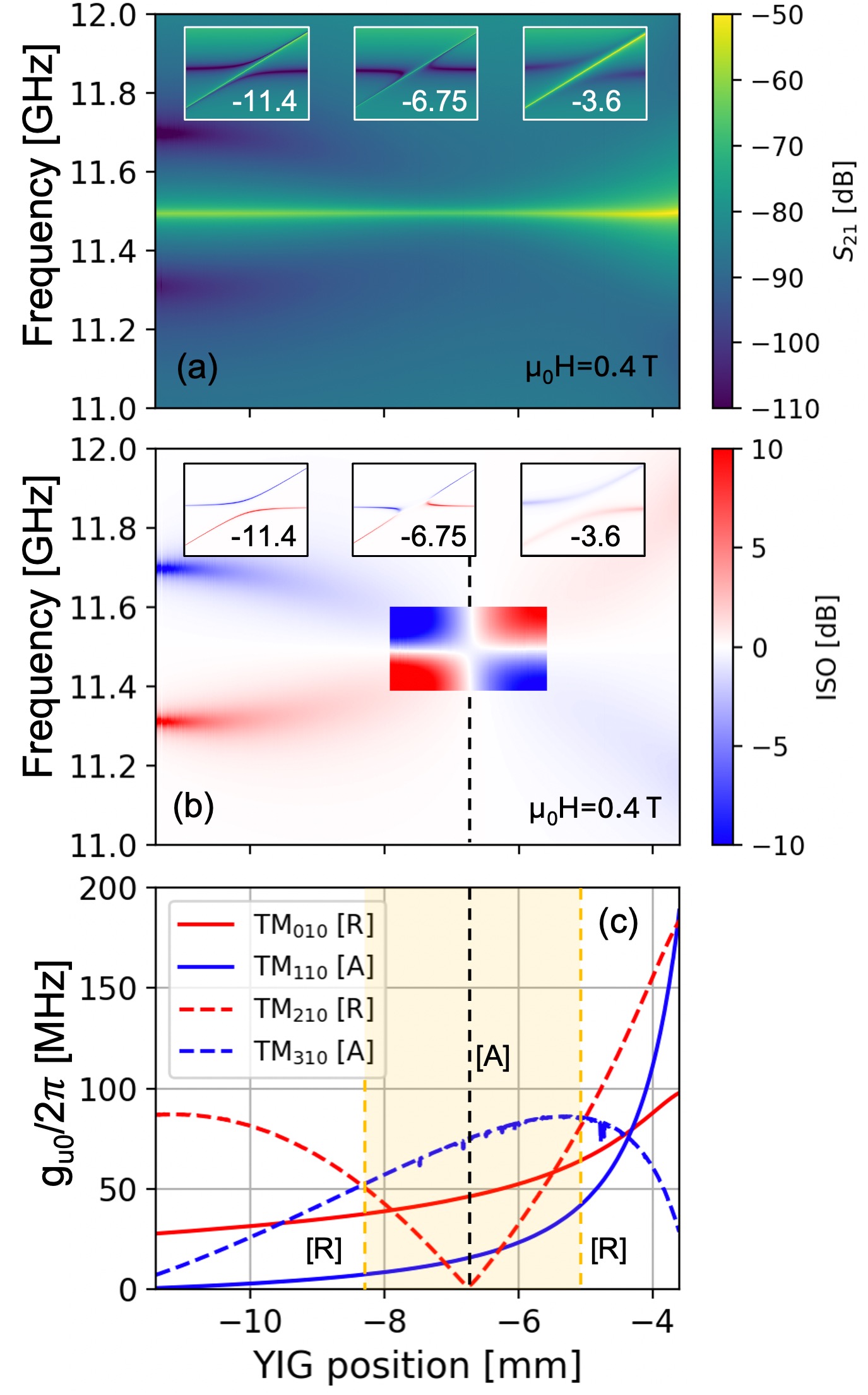}
\caption{Coupling-regime transition. (a) $S_{21}$ and (b) ISO as a function of frequency and the YIG-sphere position, scanned from $-3.6$ to $-11.4$ mm ($0$ mm corresponding to the cavity center). The calculation is performed using our model for a fixed magnetic field of $\mu_0 H = 0.4$~T (YIG resonance magnetic field). (c) Evolution of the coupling strengths $g_{u0}$ for the cavity modes. The yellow region between $-8.33$ and $-5.08$~mm corresponds to the response of an attractive regime of the system, denoted by [A], while the remaining region exhibits repulsive behavior [R]. The insets in (a) and (b) show $S_{21}$ and ISO for the YIG position located at $-11.4$, $-6.75$, and $-3.6$ mm. Note that the central inset in (b) shows the ISO variation with a scale from $-0.5$ to $0.5$ dB to enhance the contrast around the critical position.
}
\label{fig:transition}
\end{figure}

While the model successfully captures the essential physics without adjustable parameters, it also provides a robust framework to explore how the system’s behavior evolves under controlled geometric variations. To investigate the transition between level repulsion and level attraction, we systematically varied the YIG-sphere position along the axis connecting Pos.~1 and Pos.~3 in finite-element simulations. For each intermediate position, all relevant parameters are extracted following the same procedure described above. 

This approach allows us to examine in detail how the coupling nature changes through the generated antiresonance, and to study the gradual evolution from repulsive to attractive regimes. To illustrate this behavior, Fig.~\ref{fig:transition} presents the evolution of (a) $S_{21}$, (b) ISO, and (c) $g_{u0}$ as the YIG sphere is systematically displaced inside the cavity from $y = -3.6$~mm to $-11.4$~mm in steps of $0.01$~mm ($y = 0$~mm corresponding to the cavity center).

By examining the evolution of the $g_{u0}$ curves in Fig.~\ref{fig:transition}(c), one can more readily identify the critical conditions under which the repulsive and attractive coupling regimes attain their maximum contributions. As indicated in the figure, the TM modes are indexed with an [R] for repulsive and an [A] for attractive coupling, consistent with the discussion of $\Phi_{21}$ jumps introduced earlier. In the following, we consider three representative cases: the two extreme positions of the YIG sphere in the cavity ($-11.4$~mm and $-3.6$~mm), and the position corresponding to the minimum coupling to the $\mathrm{TM}_{210}$ mode ($\omega_{c2}$), indicated by the vertical black dashed line in Fig.~\ref{fig:transition}(c) at $y=-6.75$~mm.

Focusing first on the latter case, we observe that the $S_{21}$ and ISO curves calculated at $\mu_0 H = 0.4$~T, shown in Figs.~\ref{fig:transition}(a) and (b), do not exhibit any visible frequency splitting (a signature of an anticrossing) at this specific position. This observation is even more pronounced in the inset of Fig.~\ref{fig:transition}(b), where the ISO variation is shown on a restricted scale between $-0.5$ and $0.5$~dB to make the small variations more visible. Based on this visualization, one may assume that level attraction is the dominant mechanism. Indeed, when plotting both $S_{21}$ and ISO as functions of frequency and magnetic field (insets of Figs.~\ref{fig:transition}(a) and (b), indexed at $-6.75$), the response clearly exhibits an attractive character. For clarity, the cavity modes contributing to the attractive regime are $\omega_{c1}$ and $\omega_{c3}$, whereas those contributing to the repulsive regime are $\omega_{c0}$ and $\omega_{c2}$. Considering the values reported in Tab.~\ref{tab:coupling_res}, we find that at $\omega_{c3}$ ($\omega_{c1}$), the coupling strength contributing to the attractive response at the antiresonance is $g_{30}/2\pi = 75$~MHz ($g_{10}/2\pi = 15.715$~MHz), which is significantly larger than that of the $\omega_{c2}$ mode, for which $g_{20}/2\pi = 1.866$~MHz. However, the coupling to the $\omega_{c0}$ mode, with $g_{00}/2\pi = 46.171$~MHz, is not negligible compared to $g_{30}/2\pi$. Despite this, the attractive regime remains dominant and is even more pronounced than at Pos.~3. The only notable difference between the position at $-6.75$~mm and Pos.~3 lies in the value of $g_{20}/2\pi$, which increases from $1.866$~MHz to $32.403$~MHz. 

To investigate whether the couplings $g_{u0}$ contribute differently to the overall response of the system at the antiresonance, artificial values of $g_{u0}$ were tested. For instance, by setting $g_{20}/2\pi = 50$~MHz and $g_{00} = g_{10} = g_{30} = 0$, the response is purely repulsive. In this case, a coupling strength of $g_{30} = 1.136\,g_{20}$ (with $g_{00} = g_{10} = 0$) is required to suppress any coupling to $\omega_{ar}$. Similarly, a coupling strength of $g_{10} = 5.7\,g_{20}$ (with $g_{00} = g_{30} = 0$) is required to suppress the coupling. In other words, the coupling strength $g_{30}/2\pi$ contributes approximately five times more strongly to the overall response than $g_{10}/2\pi$. By applying the same procedure to suppress an attractive signature, we find that $g_{20}/2\pi$ has an impact that is about 5.2 times larger than that of $g_{00}/2\pi$.

Returning to the discussion at the position $-6.75$~mm, it now appears more clearly that the contributions of $g_{00}/2\pi$ and $g_{10}/2\pi$ to the global response at $\omega_{ar}$ are approximately five times weaker than those of the two latter modes. At the position $-11.4$~mm, the repulsive regime is maximized, as shown in the top-left insets of Figs.~\ref{fig:transition}(a) and (b) indexed at $-11.4$~mm. In contrast, the [A] modes are minimized in terms of coupling strength ($g_{10}/2\pi = 0.7$~MHz and $g_{30}/2\pi = 7.3$~MHz), whereas the [R] modes exhibit significantly larger couplings, with $g_{20}/2\pi = 86.925$~MHz and $g_{00}/2\pi = 27.5$~MHz. At the opposite extreme, $y = -3.6$~mm, a more pronounced repulsive regime is observed compared to the position at $-11.4$~mm. This behavior arises from the [R] modes, which exhibit large coupling strengths of $g_{20}/2\pi = 188.59$~MHz and $g_{00}/2\pi = 97.922$~MHz. The contribution of the [A] modes is not sufficient to counterbalance the repulsive regime, with coupling strengths of only $g_{30}/2\pi = 28.728$~MHz and $g_{10}/2\pi = 188.45$~MHz, whose effective contribution to the global response is reduced by a factor of five. The combination of our numerical code and finite-element simulations provides a clear mapping of the coupling nature, enabling us to anticipate and interpret mode-dependent interactions. 

Within the yellow region in Fig.~\ref{fig:transition}(c), defined between $-8.33$ and $-5.08$~mm, the system operates in an attractive regime [A]. Outside this range, the response becomes repulsive. The transition between [A] and [R] corresponds to a situation in which both coupling regimes cancel each other, resulting in the response of an uncoupled system. The coupling values associated with these two critical points are reported in Tab.~\ref{tab:coupling_res}. These results highlight that the nature of the magnon–antiresonance interaction can be continuously tuned via the YIG-sphere position, enabling controlled transitions between attractive and repulsive coupling regimes.

\section{\label{sec:level1}Conclusion}
We have experimentally investigated a three-dimensional cavity coupled to a 1 mm YIG sphere, demonstrating a clear transition between repulsive and attractive coupling regimes. By systematically varying the sphere's position, we mapped the evolution of the coupling strength associated with an interference-induced antiresonance and identified the critical positions separating the two regimes. The measurements were interpreted through finite-element simulations combined with an input--output model incorporating internal and external coupling phases. This theoretical framework also allowed us to extend the analysis beyond experimentally accessible configurations. Within the geometry studied, external phases govern the antiresonance features, while internal phases account for the observed nonreciprocity. A key limitation of the present configuration is that external phases cannot be independently tuned. To distinguish the respective contributions of mode number and phase terms would require a different cavity design, such as a standard cavity--magnon coupling scheme. Overall, this work demonstrates that both the nature of the coupling — whether repulsive or attractive — and its strength can be precisely controlled through the YIG sphere position.

Beyond the specific geometry studied here, the phase-resolved multi-mode framework introduced in this work opens concrete perspectives for device engineering. First, the identification of which cavity modes enhance or suppress nonreciprocity through their internal phases constitutes a design principle for on-chip microwave isolators: by selectively engineering the cavity mode spectrum and the YIG placement, one can maximize the constructive contributions to isolation while minimizing competing modes, potentially reaching isolation ratios beyond those achievable in single-mode configurations. Second, the model provides a predictive tool for cavity--magnon quantum transducers, where nonreciprocal photon routing is a key functionality. Third, the ability to continuously tune the coupling nature between repulsive and attractive regimes through a single geometric parameter — the sphere position — offers a straightforward knob for reconfigurable signal processing in hybrid magnonic architectures, including circulators and directional amplifiers. This framework therefore constitutes a versatile and predictive tool for designing tunable cavity--magnon platforms, with promising perspectives for both classical and quantum hybrid devices.

\begin{acknowledgments}

We acknowledge financial support from Brest Metropole for the PhD funding of Guillaume Bourcin. This work is part of the research program supported by the European Union through the European Regional Development Fund (ERDF), as well as by the Ministry of Higher Education and Research and the Brittany region through the CPER SpaceTech DroneTech. Jeremy Bourhill is funded by the Centre of Excellence for Dark Matter Particle Physics. We also acknowledge the financial support of the ANR project ICARUS under grant agreement ANR-22-CE24-0008-03. Finally, we thank Bernard Abiven for CNC machining of the cavities used in this study.

\end{acknowledgments}

\nocite{*}
\bibliography{aipsamp}

\newpage
\appendix

\section{Control of the antiresonance features}
\label{appendix:antires}
\begin{figure}[h!]
    \centering
\includegraphics[width=9cm]{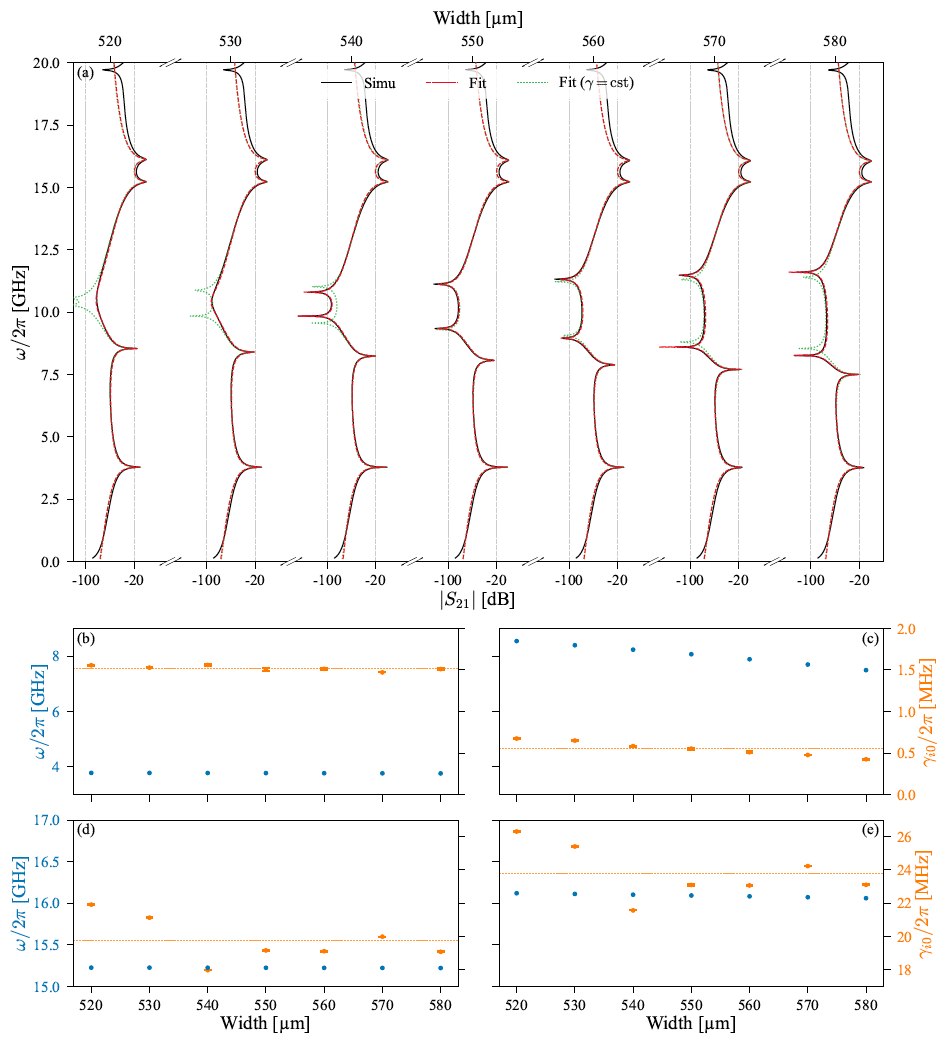}
    \caption{(a) Transmission of the dielectric-loaded cavity at different dielectric widths: FD simulation in solid black lines; input-output model with fitted $\gamma_{i0}$ values in dashed red lines; and input-output model with the mean over dielectric widths of the fitted $\gamma_{i0}$ values in dotted green line. (b)-(e) show the mode frequencies in blue and the fitted $\gamma_{i0}/2\pi$ values in orange versus the dielectric width for the first four modes. The mean values of $\gamma_{i0}/2\pi$ over the dielectric widths are represented by the dotted orange lines for each mode.}
    \label{fig:FW_FIG1a}
\end{figure}

To understand how antiresonances can occur, frequency domain (FD) simulations were performed on the cavity for eight different dielectric widths, ranging from 520 to 580 $\mu m$. The transmissions for each dielectric width are depicted as solid black lines in Fig. \ref{fig:FW_FIG1a}(a).

As shown, two antiresonances occur and repel each other in frequency as the dielectric width increases. To understand which variables cause these antiresonance dynamics, we fit each of the transmissions using the input-output model. In the FD simulations, the cavity boundaries were considered perfect electric conductors, which means that the intrinsic dissipation rates associated with the electric losses are null. The fitted variables are the external dissipation rates $\gamma_{ij}$, associated with the external coupling strengths $\kappa_{ij}$. In this cavity, two probes are considered, and their coupling to one mode is assumed to be equal, i.e., $\gamma_{i0} = \gamma_{i1}$, where $i$ represents the cavity mode considered and $0$ or $1$ represents the first or second probe in the cavity.

The fitted values of the external dissipation rates are illustrated in Fig. \ref{fig:FW_FIG1a} for (b) $\gamma_{00}$, (c) $\gamma_{10}$, (d) $\gamma_{20}$, and (e) $\gamma_{30}$ with respect to the dielectric width. Note that the mode frequencies $\omega_i$ are depicted in Fig. \ref{fig:FW_FIG1a} (b)-(e) and are known from eigenmodes simulations. The external coupling phases $\phi_{ij}$ lead to the same $\Phi_{21}$ jumps for the first and third modes, and are opposed to the second and fourth modes.

The dashed red line in Fig. \ref{fig:FW_FIG1a} (a) represents the transmission of the input-output model with the values given in Fig. \ref{fig:FW_FIG1a} (b)-(e), accurately reproducing the resonances and antiresonances in the frequency range from 3.5 to 16.5 GHz. We observe that only the frequency of the second mode depends on the dielectric width, decreasing slightly from 8.54 to 7.50 GHz as the width increases from 520 to 580 $\mu m$. Additionally, the external dissipation rates of the first mode remain relatively constant, while those of the second mode decrease. However, the external dissipation rates of the last two modes vary with respect to the dielectric width without a clear trend.

To understand the role of the second mode's frequency decrease and the variation in external dissipation rates, all transmissions were plotted as dotted green lines in Fig. \ref{fig:FW_FIG1a} (a) using the input-output model, assuming constant $\gamma$ values equal to their mean values across different dielectric widths. We also observe the attraction of the two antiresonances with decreasing dielectric width, which is primarily due to the increasing frequency of the second mode. However, without the $\gamma$ variations of the last two modes, the attraction of the two antiresonances is not sufficient to match the simulation results.

\section{Extracted coupling strength from simulations}
\label{appendix:EMandcoupling}

To investigate the coupling behavior at various YIG sphere positions, the coupling strengths for each mode were computed using eigenmodes simulations. These computed values are summarized in Tab. \ref{tab:coupling_res} and were incorporated into the input-output model alongside the dissipation rates provided in Tab. \ref{tab:fitted_gammas}.

\begin{table}[h!]
\centering
\caption{Coupling strength values and phases for different sphere positions extracted from electromagnetic simulations. R and A for level repulsion and attraction respectively. The absolute values of $\phi_{u0}$ reflect the phase reference convention adopted in the finite-element simulations. Only the relative phase difference is physically meaningful.}
\label{tab:tm_modes}
\begin{tabular}{l c c c c c}
\toprule
\textbf{Pos. [mm]}  & \textbf{F [GHz]} & \textbf{g$_{u0}$ [MHz]} & \textbf{$\phi_{u0}$ [rad]} & \textbf{$\phi_{u1}$ [rad]} & \textbf{$\theta_{u0}$ [rad]} \\
\hline
{-11.4} 
 & 3.772 & 27.872 & -1.571 & 1.571 & 0.000 \\
 & 7.775 & 0.733 & -1.571 & -1.571 & -1.552 \\
 & 15.221 & 86.925 & -1.571 & 1.571 & 3.14 \\
 & 16.076 & 7.343 & -1.571 & -1.571 & 1.183 \\
    \hline
{-10 (Pos.1)} 
 & 3.772 & 31.561 & -1.571 & 1.571 & 0.000 \\
 & 7.776 & 2.984  & -1.571 & -1.571 & -1.563 \\
 & 15.218 & 81.438 & -1.571 & 1.571 & 3.134 \\
 & 16.076 & 25.717 & -1.571 & -1.571 & 1.474 \\
    \hline
{-8.33 [R]-[A]} 
 & 3.773 & 37.469 & -1.571 & 1.571 & 0.000 \\
 & 7.776 & 7.374 & -1.571 & -1.571 & -1.571 \\
 & 15.214 & 51.703 & -1.571 & 1.571 & 3.117 \\
 & 16.076 & 51.914 & -1.571 & -1.571 & 1.540 \\
    \hline
{-8(Pos.2)} 
 & 3.771 & 38.939 & -1.571 & 1.571 & 0.000 \\
 & 7.774 & 8.626  & -1.571 & -1.571 & -1.571 \\
 & 15.213 & 42.799 & -1.571 & 1.571 & 3.108 \\
 & 16.076 & 57.073 & -1.571 & -1.571 & 1.548 \\
    \hline
{-6.75} 
 & 3.772 & 46.171 & -1.571 & 1.571 & 0.000 \\
 & 7.776 & 15.715 & -1.571 & -1.571 & -1.571 \\
 & 15.212 & 1.866  & -1.571 & 1.571 & 1.626 \\
& 16.076 & 75.036 & -1.571 & -1.571 & 1.572 \\
    \hline
{-6 (Pos.3)} 
 & 3.773 & 52.479 & -1.571 & 1.571 & 0.000 \\
 & 7.776 & 23.345 & -1.571 & -1.571 & -1.571 \\
 & 15.212 & 32.403 & -1.571 & 1.571 & 0.064 \\
 & 16.076 & 82.726 & -1.571 & -1.571 & 1.584 \\
  \hline
{-5.08 [A]-[R]} 
 & 3.773 & 64.055 & -1.571 & 1.571 & 0.000 \\
 & 7.776 & 41.691 & -1.571 & -1.571 & -1.571 \\
 & 15.214 & 80.723 & -1.571 & 1.571 & 0.027 \\
 & 16.076 & 85.71 & -1.571 & -1.571 & 1.600 \\
  \hline
 {-3.6} 
 & 3.771 & 97.922 & -1.571 & 1.571 & 0.000 \\
 & 7.774 & 188.45 & -1.571 & -1.571 & -1.571 \\
 & 15.219 & 183.59 & -1.571 & 1.571 & 0.004 \\
 & 16.076 & 28.728 & -1.571 & -1.571 & 1.759 \\
\label{tab:coupling_res}
\end{tabular}
\end{table}

\section{Transmission spectra for all YIG positions}
\label{appendix:YIGposmeas}

\begin{figure}[h!]
    \centering
    \includegraphics[width=9cm]{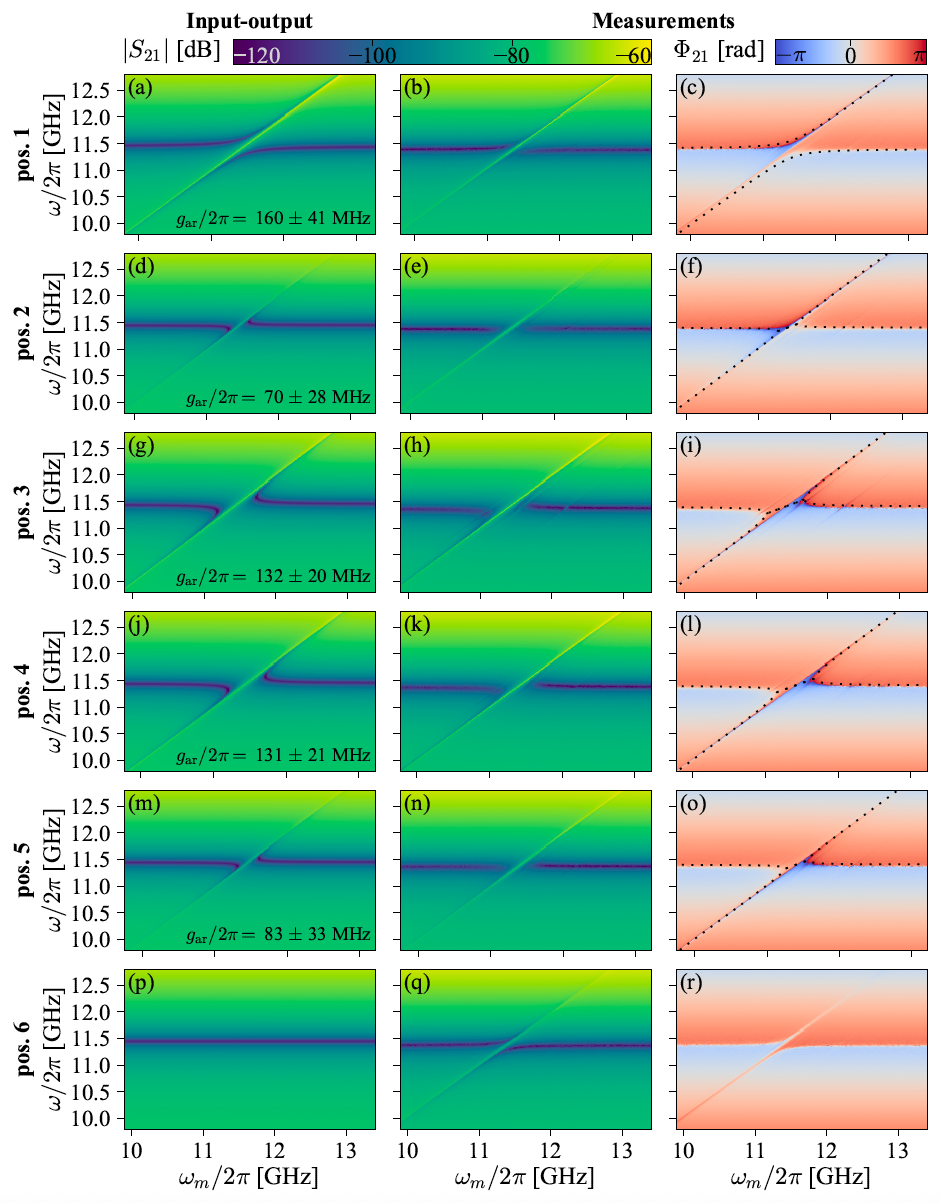}
    \caption{Transmission spectra at the 6 YIG sphere positions, showing the magnitude for the input-output model in the first column, the measurements in the second column, and the phase of the measurements in the third column. The antiresonance polariton frequencies from the input-output transmission spectra are indicated by dotted black lines in the third column.}
    \label{fig:FW_FIG3a}
\end{figure}

The transmission spectra from the input-output model for all YIG positions are illustrated in the first column of Fig. \ref{fig:FW_FIG3a}. The Gilbert damping rate was arbitrarily set to $\alpha = 4.10^{-4}$ to align with the measurement observations. Note that no fitting was performed on $\alpha$ as it does not affect the coupling behavior. The increase in damping rate could be attributed to the excitation of multiple magnon modes, driven by the inhomogeneous RF magnetic field concentrated on the YIG.

These spectra align with the previous interpretation of the modes' field distributions. Specifically, at position 1, the antiresonance coupling exhibits level repulsion, whereas at position 3, it shows level attraction. Furthermore, position 6 also demonstrates no coupling, meaning that the coupling effectively decreased by compensation between attractive and repulsive modes.

The second column of Fig. \ref{fig:FW_FIG3a} presents the measured transmission spectra for all YIG positions. It is evident that interpreting the antiresonance coupling behavior based on the field distribution at the antiresonance frequency did not accurately predict the coupling behavior. In particular, it was anticipated that position 1 would exhibit a weaker repulsive coupling compared to position 6; however, the measurements reveal the opposite trend.

The third column in Fig. \ref{fig:FW_FIG3a} depicts the measured phases $\Phi_{21}$ for all YIG positions. Although the coupling behavior is not always clear from certain transmission measurements - where coupling with higher magnon modes obscures the antiresonance polaritons around the coupling frame - the $\Phi_{21}$ jump in the phase spectra helps track the polariton frequencies. The dotted black line in the third column represents the fitted antiresonance polaritons according to the input-output model.

\begin{figure}[h!]
    \centering
    \includegraphics[width=9cm]{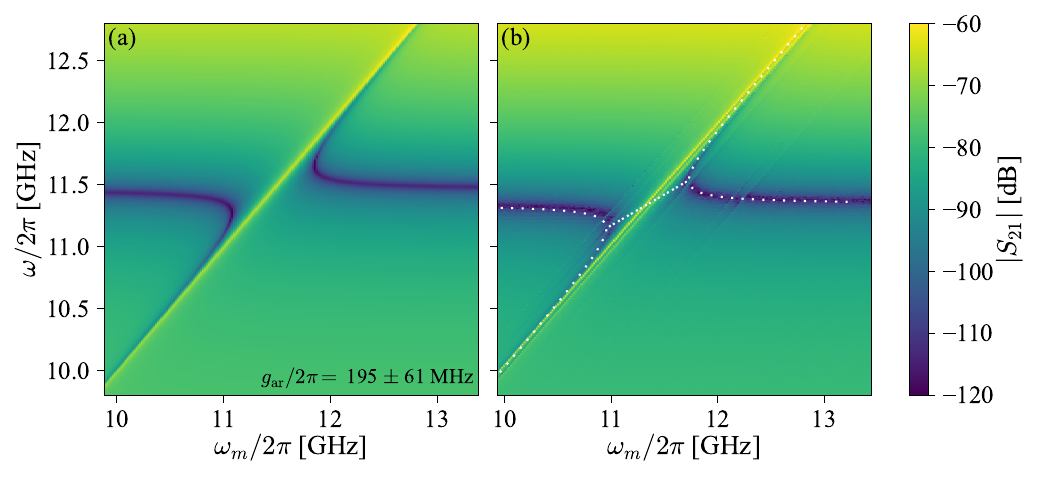}
    \caption{Transmission spectra with two YIG spheres positioned at 3 and 3b for (a) the input-output model and (b) measurements. In (b), the antiresonance polariton frequencies from the input-output model are indicated by a dotted white line.}
    \label{fig:FW_FIG4a}
\end{figure}

We observe that the evolution of the antiresonance coupling across different YIG positions is consistent between the input-output model and the measurements. Specifically, positions 1, 3, 4, and 5 exhibit the same coupling strength and nature. However, for positions 2 and 6, where the coupling strengths are respectively very low or does not occur, the coupling behavior differs, and the measurements indicate level repulsion. This discrepancy can be attributed to the fact that the resonance coupling strengths were extracted from EM simulations, which used different mode frequencies than those measured. Consequently, the coupling strength may vary slightly, leading to different antiresonance coupling behaviors at low coupling strengths.

In Fig. \ref{fig:FW_FIG4a}, we present the transmission spectra of (a) the input-output model and (b) the measurement with two YIG spheres placed at positions 3 and 3b. For the model, the coupling strengths of each mode at position 3b are assumed to be the same as those at position 3, based on symmetry considerations of the cavity. The effective antiresonance coupling strength in the model is $g_\mathrm{ar}/2\pi = 195 \pm 61$ MHz, and the antiresonance polariton frequencies fit well with the measurements, as indicated by the dotted white line in Fig. \ref{fig:FW_FIG4a}(b). This demonstrates that by considering the field distribution of the cavity modes, we can tune the antiresonance as desired and increase the coupling strength, regardless of the coupling nature.

\end{document}